\documentclass[reprint,floatfix,amsfonts, amsmath, showpacs,aps,superscriptaddress]{revtex4-1}
\usepackage{bm}
\usepackage{graphicx}
\usepackage{epstopdf}

\begin{document}
\title{Muon spin relaxation and inelastic neutron scattering investigations of \\ all-in/all-out antiferromagnet \bm{${\rm Nd_2Hf_2O_7}$}}
\author{V. K. Anand}
\altaffiliation{vivekkranand@gmail.com}
\affiliation{\mbox{Helmholtz-Zentrum Berlin f\"{u}r Materialien und Energie GmbH, Hahn-Meitner Platz 1, D-14109 Berlin, Germany}}
\author{D. L. Abernathy}
\affiliation{\mbox{Quantum Condensed Matter Division, Neutron Sciences Directorate, Oak Ridge National Laboratory, Oak Ridge} Tennessee 37831, USA}
\author{D. T. Adroja}
\affiliation{ISIS Facility, Rutherford Appleton Laboratory, Chilton, Didcot, Oxon, OX11 0QX, United Kingdom}
\affiliation{\mbox{Highly Correlated Matter Research Group, Physics Department, University of Johannesburg, P.O. Box 524,} Auckland Park 2006, South Africa}
\author{A. D. Hillier}
\author{P. K. Biswas}
\affiliation{ISIS Facility, Rutherford Appleton Laboratory, Chilton, Didcot, Oxon, OX11 0QX, United Kingdom}
\author{B. Lake}
\altaffiliation{bella.lake@helmholtz-berlin.de}
\affiliation{\mbox{Helmholtz-Zentrum Berlin f\"{u}r Materialien und Energie GmbH, Hahn-Meitner Platz 1, D-14109 Berlin, Germany}}

\date{\today}

\begin{abstract}
Nd$_2$Hf$_2$O$_7$, belonging to the family of geometrically frustrated cubic rare earth pyrochlore oxides, was recently identified to order antiferromagnetically below $T_{\rm N}\approx 0.55$~K with an all-in/all-out arrangement of Nd$^{3+}$ moments, however with a much reduced ordered state moment. Herein we investigate the spin dynamics and crystal field states of Nd$_2$Hf$_2$O$_7$ using muon spin relaxation ($\mu$SR) and inelastic neutron scattering (INS) measurements. Our $\mu$SR  study confirms the long range magnetic ordering and shows evidence for coexisting persistent dynamic spin fluctuations deep inside the ordered state down to 42~mK. The INS data show the crytal electric field (CEF) excitations due to the transitions both within the ground state multiplet and to the first excited state multiplet. The INS data are analyzed by a model based on CEF and crystal field states are determined. Strong Ising-type anisotropy is inferred from the ground state wavefunction. The CEF parameters indicate the CEF-split Kramers doublet ground state of Nd$^{3+}$ to be consistent with the dipolar-octupolar character. 
\end{abstract}

\maketitle

\section{\label{Intro} INTRODUCTION}

The rare earth pyrochlore oxides $R_2B_2$O$_7$ ($R$ is a trivalent rare earth ion and $B$ a tetravalent transition metal ion or Ge, Sn, Pb) consisting of corner sharing tetrahedra of R$^{3+}$ ions present diverse emergent magnetic states due to the interplay of crystal electric field (CEF), antiferromagnetic exchange and ferromagnetic dipolar interactions \cite{Gardner2010, Castelnovo2012, Gingras2014, Ramirez1999, Siddharthan1999,Bramwell2001} and form an interesting topic of current research activities in condensed matter physics. The crystal topology and CEF play an important role in determining the ground state properties of these pyrochlores, leading to an Ising-type anisotropy in spin-ice materials  Dy$_2$Ti$_2$O$_7$ and Ho$_2$Ti$_2$O$_7$ \cite{Ramirez1999, Siddharthan1999, Hertog2000, Bramwell2001}. With a local $\langle 111 \rangle $ anisotropy and ferromagnetic interactions, the ground state of these classical spin-ice materials correspond to `two-in/two-out' spin configuration which possesses a Pauling residual entropy of $(1/2) R \ln(3/2)$ \cite{Harris1997,Ramirez1999}. The fundametal excitation of a spin ice material is the magnetic monopole \cite{Castelnovo2008} which is quite striking. Interestingly, among the diverse magnetic state of these 227 pyrochlores, the antiferromagnetic (AFM) all-in/all-out (AIAO) state has recently been predicted to show interesting physics by introducing the concepts of `double monopoles' and `staggered charge fluid and crystal' \cite{Bartlett2014,Guruciaga2014}.  

The phase diagram of an Ising pyrochlore (dipolar model) incorporates both spin-ice and antiferromagnetic AIAO states depending on the relative strengths of antiferromagnetic exchange and ferromagnetic dipolar interaction \cite{Hertog2000,Melko2004}. The monopole dynamics is believed to provide the key to the understanding of dipolar spin-ice \cite{Castelnovo2008,Morris2009,Jaubert2009,Bramwell2009,Fennell2009}. The magnetic monopoles emerge as quasiparticle excitations in the Coulomb phase corresponding to `three-in/one-out' or `one-in/three-out' configurations and interact via magnetic Coulomb potential \cite{Castelnovo2008}. Recent theoretical works propose that the `all-in' and `all-out' spin configurations also correspond to excitations in the Coulomb phase, referred as `double monopoles' as they involve `double excitations' \cite{Bartlett2014,Guruciaga2014}.  The AIAO AFM systems are thus considered important for understanding the monopole dynamics in spin ice materials.

AIAO AFM ordering has recently been found in a few of the 227 compounds including Eu$_2$Ir$_2$O$_7$ \cite{Zhao2011, Sagayama2013}, Nd$_2$Ir$_2$O$_7$ \cite{Tomiyasu2012,Guo2013}, Nd$_2$Sn$_2$O$_7$ \cite{Bertin2015}, Nd$_2$Hf$_2$O$_7$ \cite{Anand2015} and Nd$_2$Zr$_2$O$_7$ \cite{Lhotel2015,Xu2015}. While in Eu$_2$Ir$_2$O$_7$ only Ir$^{4+}$ orders, in Nd$_2$Ir$_2$O$_7$ both Nd$^{3+}$ and Ir$^{4+}$ moments order with the AIAO spin arrangement \cite{Tomiyasu2012,Guo2013}. On the other hand in Nd$_2$Sn$_2$O$_7$, Nd$_2$Hf$_2$O$_7$ and Nd$_2$Zr$_2$O$_7$ the $B$ site is nonmagnetic and only Nd$^{3+}$ moments order. Among these AIAO ordered systems, the compounds containing Nd$^{3+}$ moments with total angular momenta $J = 9/2$ are of particular interest because of the possible exotic behavior due to the dipolar-octupolar character of Kramers doublet ground state of Nd$^{3+}$. The theoretical treatment by Huang {\it et al}.\ \cite{Huang2014} suggests that the Kramers doublet ground state of rare earths with $J = 9/2$ (Nd$^{3+}$) and $J = 15/2$ (Dy$^{3+}$) under space group symmetry may transform with octupolar component in addition to the dipolar term. While the $x$ and $z$ components of time-reversal odd pseudospin operator transform like a magnetic dipole, the $y$ component transforms as a component of the magnetic octupole tensor, accordingly systems with dipolar-octopolar Kramers doublet ground state are predicted to show two distinct quantum spin-ice (QSI) phases dubbed as dipolar QSI and octupolar QSI \cite{Huang2014}. This adds another attribute to the Nd-based AIAO AFM compounds on account of dipolar-octopolar nature of Kramers doublet ground state of Nd$^{3+}$. 

Nd$_2$Sn$_2$O$_7$ is found to order antiferromagnetically below $T_{\rm N }\approx 0.91$~K for which powder neutron diffraction (ND) has revealed an AIAO magnetic structure with an ordered moment of 1.708(3)~$\mu_{\rm B}$ at 0.06~K \cite{Bertin2015}. Further, evidence for persistent spin dynamics in the ordered state of Nd$_2$Sn$_2$O$_7$ is found from the muon spin relaxation ($\mu$SR) study which also detects anomalously slow spin dynamics in the paramagnetic state up to $\sim 30\,T_{\rm N }$ \cite{Bertin2015}. Nd$_2$Zr$_2$O$_7$ is found to order antiferromagnetically below $T_{\rm N }\approx 0.4$~K with an ordered moment of 1.26(2)~$\mu_{\rm B}$/Nd at 0.1~K \cite{Xu2015} and 0.80(5)~$\mu_{\rm B}$/Nd at 0.15~K \cite{Lhotel2015} in AIAO state, determined by powder ND measurements. The $\mu$SR study on Nd$_2$Zr$_2$O$_7$ showed no clear signature of long range ordering, instead persistent spin dynamics is inferred \cite{Xu2016}. A recent inelastic neutron scattering study reports observation of magnetic spin fragmentation in Nd$_2$Zr$_2$O$_7$ signifying possible spin-ice behavior in this compound \cite{Petit2016}  

Very recently we investigated the magnetic properties of the pyrohafnate Nd$_2$Hf$_2$O$_7$ \cite{Anand2015} and found evidence for long-range antiferromagnetic ordering below $T_{\rm N }\approx 0.55$~K\@.  Strong local $\langle 111 \rangle$ Ising anisotropy was evidenced from the magnetic data which are well described by an effective pseudo spin-half model. The magnetic structure was determined by neutron powder diffraction which revealed an all-in/all-out arrangement of Nd$^{3+}$ moments characterized by propagation wave vector {\bf k} = (0,\,0,\,0). However, the ordered magnetic moment was found to be only 0.62(1)~$\mu_{\rm B}$/Nd at 0.1~K \cite{Anand2015}, much lower than the expected 2.5 $\mu_{\rm B}$/Nd for the Ising ground state of this compound. Such a strong reduction of moment reflects the presence of strong quantum fluctuations in the ordered state possibly due to the octupolar coupling of the Kramers doublet of Nd$^{3+}$. A recent theoretical work by Guruciaga {\it et al}.\ proposes thermal order by disorder that can be tuned by field in antiferromagnetic Ising pyrochlores like Nd$_2$Zr$_2$O$_7$  and Nd$_2$Hf$_2$O$_7$ \cite{Guruciaga2016}.

Extending our work on Nd$_2$Hf$_2$O$_7$, in order to probe the spin dynamics we have carried out the $\mu$SR measurements on Nd$_2$Hf$_2$O$_7$. We also performed INS measurements to determine the crystal field states and check the dipolar-octupolar nature of CEF-split Kramers doublet ground state of Nd$_2$Hf$_2$O$_7$. The long range magnetic ordering is confirmed by $\mu$SR. In addition we also see evidence for persistent dynamical fluctuations in the orderered state. The CEF excitations are clearly seen in INS data, the analysis of which prove that the wavefunction of the Kramers doublet ground state of Nd$^{3+}$ ($J=9/2$) is compatible with a dipolar-octupolar type behavior as anticipated in view of strongly reduced ordered state moment \cite{Anand2015}.

\section{\label{ExpDetails} EXPERIMENTAL DETAILS}

Polycrystalline samples of Nd$_2$Hf$_2$O$_7$ and La$_2$Hf$_2$O$_7$ were prepared by solid state reaction method using the stoichiometric mixture of high purity materials Nd$_2$O$_3$ (99.99\%) or La$_2$O$_3$ (99.999\%) and HfO$_2$ (99.95\%) as detailed in Ref.~\cite{Anand2015}. The quality of the samples were checked by room temperature powder x-ray diffraction  which revealed the single phase nature of both samples.  

Muon spin relaxation measurements were carried out at the ISIS facility, Rutherford Appleton Laboratory, Didcot, U.K. using the MuSR spectrometer both in zero field (ZF) and in longitudinal fields (LF) up to 0.3~T\@.  For these measurements the powder sample of Nd$_2$Hf$_2$O$_7$ was mounted on a high purity silver plate using diluted GE varnish (covered with a thin silver foil). Temperatures down to 42~mK was acheived using a dilution refrigerator. The ZF $\mu$SR spectra were recorded at several temperatures between 42~mK to 3.5~K, and LF data were collected at 0.17~K and 1.0~K for fields between 5~mT to 0.3~T\@. 

The inelastic neutron scattering measurements were carried out at the Spallation Neutron Source (SNS), Oak Ridge National Laboratory (ORNL), USA using the direct geometry time-of-flight spectrometer ARCS \cite{ARCS}.  For these measurements about 20~g samples each of Nd$_2$Hf$_2$O$_7$ and La$_2$Hf$_2$O$_7$ were mounted inside thin double-walled cylindrical aluminium cans casting the powdered samples in the form of cylinderical annuli. In order to access the full range of excitations the INS responses were collected at 5~K and 300~K using neutrons of incident energies $E_i = 50$~meV, 150~meV, 400~meV and 700~meV.

\section{\label{Sec:muSR} Muon Spin Relaxation}

\begin{figure}
\includegraphics[width=\columnwidth]{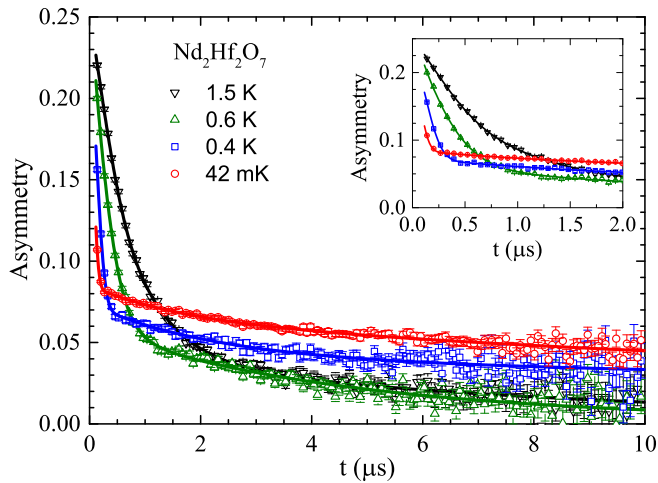}
\caption{\label{fig:MuSR1} (Color online) Zero field muon spin asymmetry function $G_z$ versus time $t$ spectra of Nd$_2$Hf$_2$O$_7$ at few representative temperatures. Solid curves are the fits to the $\mu$SR data by the relaxation function in Eq.~(\ref{eq:muSR}). Inset: An expanded plot for $t\leq 2~\mu$s. }
\end{figure}

\begin{figure}
\includegraphics[width=\columnwidth]{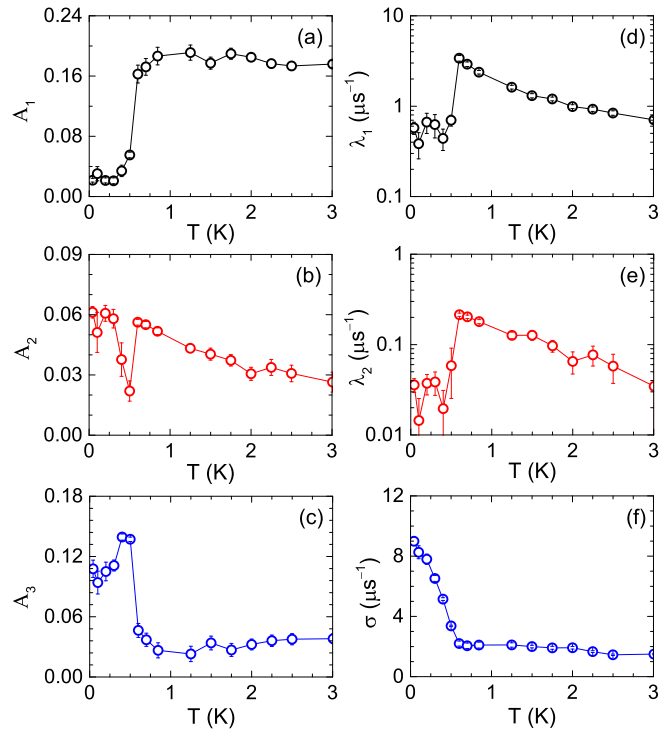}
\caption{\label{fig:MuSR2} (Color online) Temperature $T$ dependence of the initial asymmetries (a) $A_1$, (b) $A_2$, (c) $A_3$, and the depolarization rates (d) $\lambda_1$, (e) $\lambda_2$, (f)  $\sigma$, obtained from the analysis of the zero field $\mu$SR data of Nd$_2$Hf$_2$O$_7$ at $42~{\rm mK} \leq T \leq 3$~K\@.}
\end{figure}

In order to probe the spin dynamics of Nd$_2$Hf$_2$O$_7$ we carried out $\mu$SR measurements in zero field (ZF) as well as for magnetic field applied in longitudinal geometry, i.e., along the initial spin direction of the muon. The ZF $\mu$SR asymmetry spectra ($G_z$ as a function of time $t$) for few representative temperatures between 42~mK and 1.5~K are shown in Fig.~\ref{fig:MuSR1}. A clear change in $\mu$SR spectra is evident at temperatures above and below $T_{\rm N} \approx 0.55$~K\@. Even though we do not see any clear oscillation (related to muon spin precession about a well defined magnetic field) in the $\mu$SR spectra in the ordered state, a loss in initial asymmetry clearly distinguishes the antiferromagnetically ordered state from the paramagnetic state. Further, we notice that at $T < T_{\rm N}$ initially at low $t$ the initial asymmetry decreases rapidly, with a slower rate of decrease at higher $t$\@. This kind of sharp decrease in asymmetry at short times has recently been observed in the ordered state of the all-in/all-out antiferromagnet Nd$_2$Sn$_2$O$_7$, though in that case oscillations in the time dependent asymmetry were also observed \cite{Bertin2015}. The oscillations were also  observed  in the $\mu$SR spectra of all-in/all-out antiferromagnet Nd$_2$Ir$_2$O$_7$ \cite{Disseler2012, Guo2013}. On the other hand no such oscillations were observed in the $\mu$SR spectra of Nd$_2$Zr$_2$O$_7$ which also has all-in/all-out antiferromagnetic order \cite{Xu2016}. Moreover no oscillating asymmetry was observed in the ordered state $\mu$SR spectra of Tb$_2$Sn$_2$O$_7$ \cite{Reotier2012, Bert2006}, Er$_2$Ti$_2$O$_7$ \cite{Lago2005, Reotier2012} or Yb$_2$Sn$_2$O$_7$ \cite{Yaouanc2013, Lago2014}. We would also like to point out that the present data were collected at a pulsed muon source which does not have enough resolution at very short time to observe the strongly damped oscillations (in very short time window) in the asymmetry.

The $\mu$SR spectra could be successfully analyzed using a relaxation function consisting of a combination of two Lorentzian and one Gaussian terms, accounting for both static and dynamic local fields at muon sites,
\begin{equation}
\begin{split}
 G_z(t) = & \ A_{1} \exp({-\lambda_1 t}) + A_{2} \exp({-\lambda_2 t})\\
             & + A_{3} \exp\left(-\frac{\sigma^2 t^2}{2}\right)+ A_{\rm BG},
\end{split}
\label{eq:muSR}
\end{equation}
where $A_{1}$, $A_{2}$ and $A_{3}$ are the initial asymmetries of the three components, and $\lambda_1$, $\lambda_2$ and $\sigma$ are the depolarization rates. The first two exponential terms (Lorentzian form) in Eq.~(\ref{eq:muSR}) account for the dynamic magnetic fluctuations (fast and slow relaxation components, $\lambda_1 > \lambda_2$) and the third term (Gaussian form) accounts for an isotropic Gaussian distribution of static fields. The last term $A_{\rm BG}$ is a constant background accounting for the muons stopping on the silver sample holder. The $A_{\rm BG} = 0.01$ was estimated from fitting the  spectra at 3~K which was then fixed for all other temperatures.

The three relaxation terms in Eq. (1) suggests three possible muon stopping sites in Nd$_2$Hf$_2$O$_7$. In a recent study using DFT calculations Foronda et al.\ \cite{Foronda2015} have identified the three possible muon stopping sites in Pr$_2T_2$O$_7$ ($T=$ Sn, Zr, Hf). The $\mu$SR data of Yb$_2$Ti$_2$O$_7$ \cite{Chang2014} were analyzed using two Lorentzian relaxation components like the present compound, however no Gaussian term was required. For the present compound while the $\mu$SR data at $T> T_{\rm N}$ are well described by two Lorentzian components, the $\mu$SR data at $T< T_{\rm N}$ need an additional Gaussian term. In order to avoid any abrupt change in fit parameters as a result of the change in fitting function at temperatures above and below $T_{\rm N}$ we analyze $\mu$SR data with relaxation function in Eq.~(\ref{eq:muSR}) in the whole temperature range. As can be seen from the fit parameters the Gaussian contribution above $T_{\rm N}$ is very small and almost $T$-independent.

The fits of the $\mu$SR spectra by the combination of Lorenztian and Gaussian decays in Eq.~(\ref{eq:muSR}) are shown by the solid curves in Fig.~\ref{fig:MuSR1}. The fit parameters obtained are shown in Fig.~\ref{fig:MuSR2}  as a function of temperature. A transition at $T_{\rm N }\approx 0.55$~K is quite clear from the $T$ dependences of parameters $\lambda_1$, $\lambda_2$ and $\sigma$ [Figs.~\ref{fig:MuSR2}(d)--(f)]. While both $\lambda_1$ and $\lambda_2$ decreases very rapidly below $T_{\rm N }$,  $\sigma$ increases. This suggests that the Gaussian contribution (static field) grows below $T_{\rm N }$ and reflects a slowing down of spin fluctuations. The static field at muon sites increases due to an increase in ordered moment as $T$ is lowered. The asymmetry of the Gaussian component ($A_3$) shows a peak near $T_{\rm N }$. While the asymmetry $A_1$ of the fast relaxing dynamic component shows a sharp decrease near $T_{\rm N }$ and a nearly $T$-independent behavior in ordered state, the asymmetry $A_2$ of the slow relaxing dynamic fluctuation drops rapidly initially but eventually increases below 0.5~K\@.  The increase of $A_2$ is accompanied by a decrease in $A_3$. Possibly, the muons sense a distribution of local fields.

\begin{figure}
\includegraphics[width=\columnwidth]{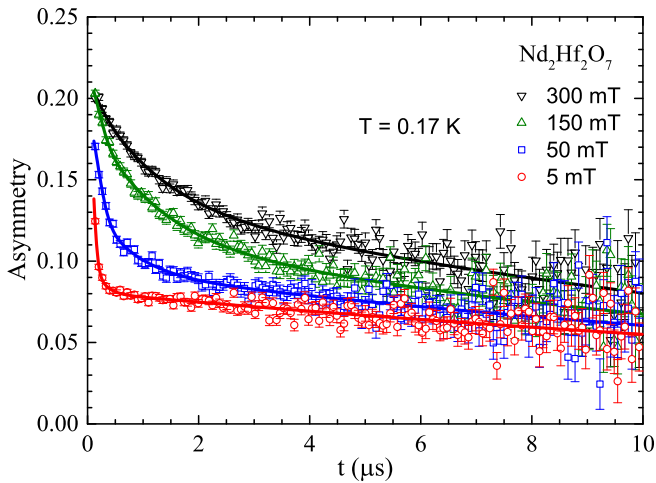}
\caption{\label{fig:MuSR3} (Color online) Longitudinal field muon spin asymmetry function $G_z$ versus time $t$ spectra of Nd$_2$Hf$_2$O$_7$ at few representative fields at 0.17~K\@. Solid curves are the fits to the $\mu$SR data by the relaxation function in Eq.~(\ref{eq:muSR}).}
\end{figure}

The sharp decrease in $\lambda_1(T)$ and $\lambda_2(T)$ below the maxima near $T_{\rm N }$ can be understood to be the result of the slowing-down of critical fluctuations at the antiferromagnetic transition. However, in the ordered state at $T\ll T_{\rm N }$ one expects a vanishing $\lambda$. Contrary to such an expectation we notice that both $\lambda_1$ and $\lambda_2$ shows a nonvanising plateau in the limit of $T\rightarrow 0$ indicating the presence of dynamic fluctuations down to 42~mK. A similar plateau was recently observed in the $\lambda(T)$ of Nd$_2$Sn$_2$O$_7$ \cite{Bertin2015} which was interpreted to be the signature of persistent spin dynamics. We thus see that the muons, in addition to long range ordering, also show dynamic spin fluctuations deep inside the ordered state of Nd$_2$Hf$_2$O$_7$.

\begin{figure}
\includegraphics[width=\columnwidth]{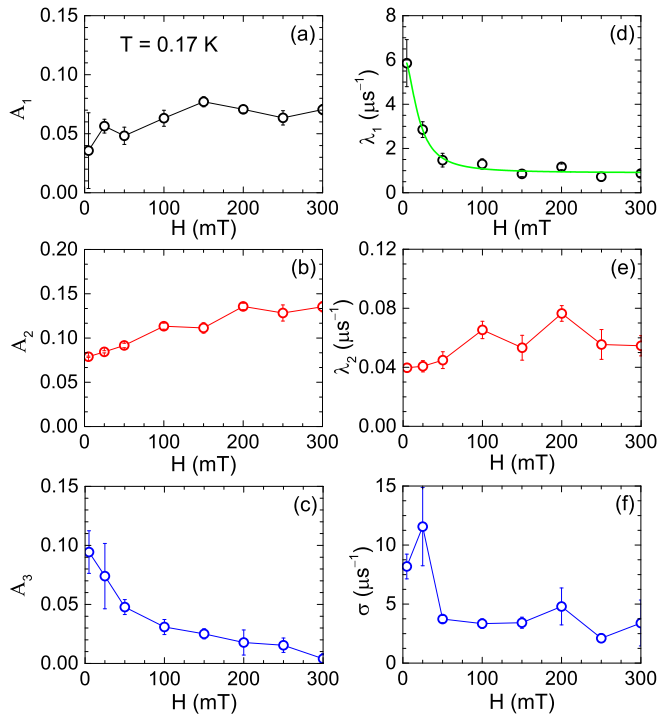}
\caption{\label{fig:MuSR4} (Color online) Magnetic field $H$ dependence of the initial asymmetries (a) $A_1$, (b) $A_2$, (c) $A_3$, and the depolarization rates (d) $\lambda_1$, (e) $\lambda_2$, (f)  $\sigma$, obtained from the analysis of the longitudinal field $\mu$SR data of Nd$_2$Hf$_2$O$_7$ at $T= 0.17$~K and $50~{\rm mT} \leq H \leq 300$~mT\@. }
\end{figure}

The LF $\mu$SR asymmetry spectra measured at 0.17~K are shown in Fig.~\ref{fig:MuSR3} for few representative fields between 5~mT and 300~mT. From the raw data it is seen that the initial asymmetry increases with increasing field. The LF $\mu$SR spectra were also analyzed by Eq.~(\ref{eq:muSR}). The fit parameters are shown in Fig.~\ref{fig:MuSR4}. It is seen that the value of $\lambda_1$ at 5.0~mT [Fig.~\ref{fig:MuSR4}(d)] drastically increases compared to its value in zero field [Fig.~\ref{fig:MuSR2}(d)]. No such increase is observed in the values of $\lambda_2$ or $\sigma$ at 5.0~mT. With further increasing field $\lambda_1$  decreases rapidly initially and then shows a weak field dependence at $H\geq 50$~mT [Fig.~\ref{fig:MuSR4}(d)]. The $\sigma(H)$ [Fig.~\ref{fig:MuSR4}(f)] increases initially and then at $H\geq 50$~mT remains nearly $H$-independent (with a reduced value).  The $\lambda_2(H)$ on the other shows a weak increase with field [Fig.~\ref{fig:MuSR4}(e)]. A weak increase is also seen in asymmetries $A_1$ [Fig.~\ref{fig:MuSR4}(a)] and $A_2$ [Fig.~\ref{fig:MuSR4}(b)], whereas $A_3$ decreases with increasing field [Fig.~\ref{fig:MuSR4}(c)]. 

We use $\lambda_1(H)$ data in Fig.~\ref{fig:MuSR4}(d) to estimate the spin autocorrelation time $\tau_c$ of spin fluctuation using the Redfield equation, 
\begin{equation}
 \lambda(H) = \lambda_0+ \frac{2\gamma_\mu^2 \langle H_{\rm loc}^2 \rangle \tau_c}{1+\gamma_\mu^2 H^2 \tau_c^2}
\label{eq:Lambda_tau}
\end{equation}
where $\lambda_0$ is $H$-independent depolarization rate,  $\gamma_\mu$ is the muon gyromagnetic ratio and $\langle H_{\rm loc}^2 \rangle $ is the time average of the second moment of the time-varying local field $H_{\rm loc}(t)$ at muon sites due to the fluctuations of neighboring Nd 4f moments. The fit of $\lambda(H)$ data by Eq.~(\ref{eq:Lambda_tau}) is shown by solid green curve in Fig.~\ref{fig:MuSR4}(d). A good fit is obtained with the fitting parameters $\lambda_0=0.90(8)~\mu$s$^{-1}$, $\surd{\langle H_{\rm loc}^2 \rangle} = 2.4(2)$~mT, and $\tau_c = 6.2(8)\times 10^{-7}$~s\@. We thus find a correlation time of about 600~ns for spin fluctations in  Nd$_2$Hf$_2$O$_7$. A correlation time of about 100~ns was found for spin fluctations in  Nd$_2$Sn$_2$O$_7$ \cite{Bertin2015}.
  
\section{\label{INS} Inelastic Neutron Scattering and Crystal Field Excitations}

\begin{figure*}
\includegraphics[width=\textwidth, keepaspectratio]{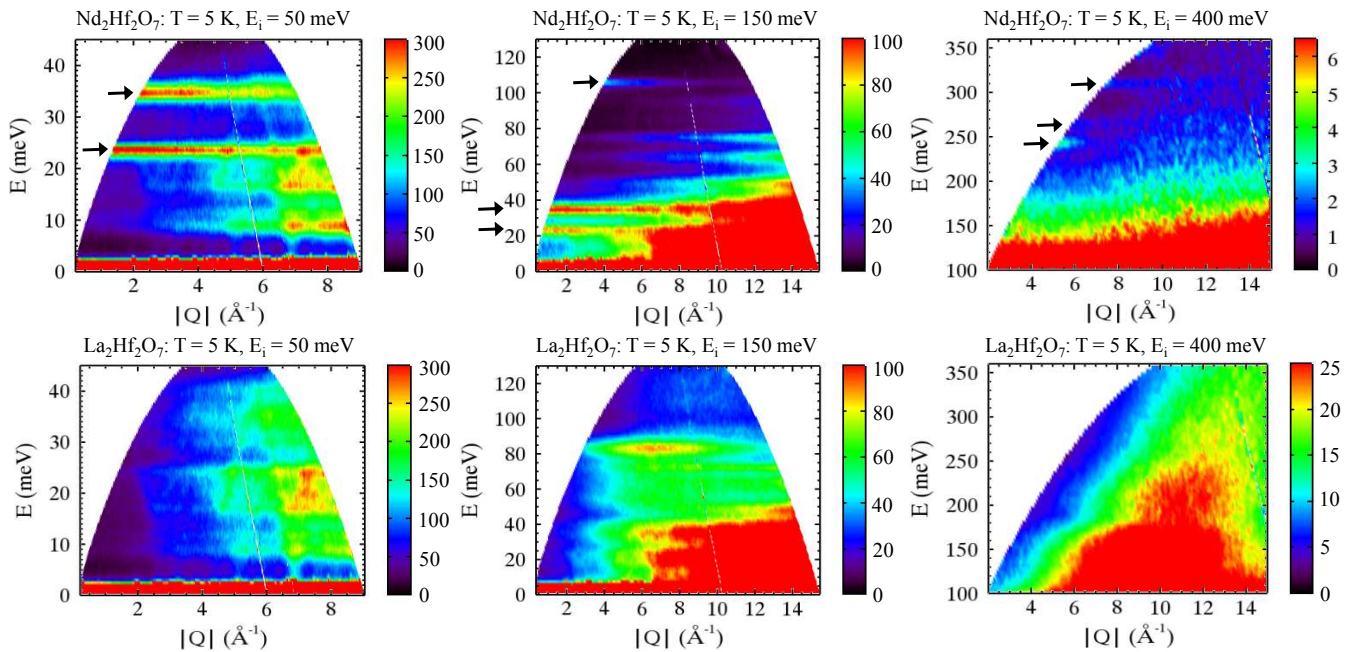}
\caption {(color online) Inelastic neutron scattering response, a color-coded map of the intensity, energy transfer ($E$) versus momentum transfer ($Q$) for Nd$_2$Hf$_2$O$_7$ (upper panels)  and La$_2$Hf$_2$O$_7$ (lower panels) measured at 5~K with the incident energies $E_i=50$~meV, 150~meV and 400~meV. The arrows mark the crystal field excitations in Nd$_2$Hf$_2$O$_7$.}
\label{fig:INS1}
\end{figure*}

\begin{figure}
\includegraphics[width=\columnwidth, keepaspectratio]{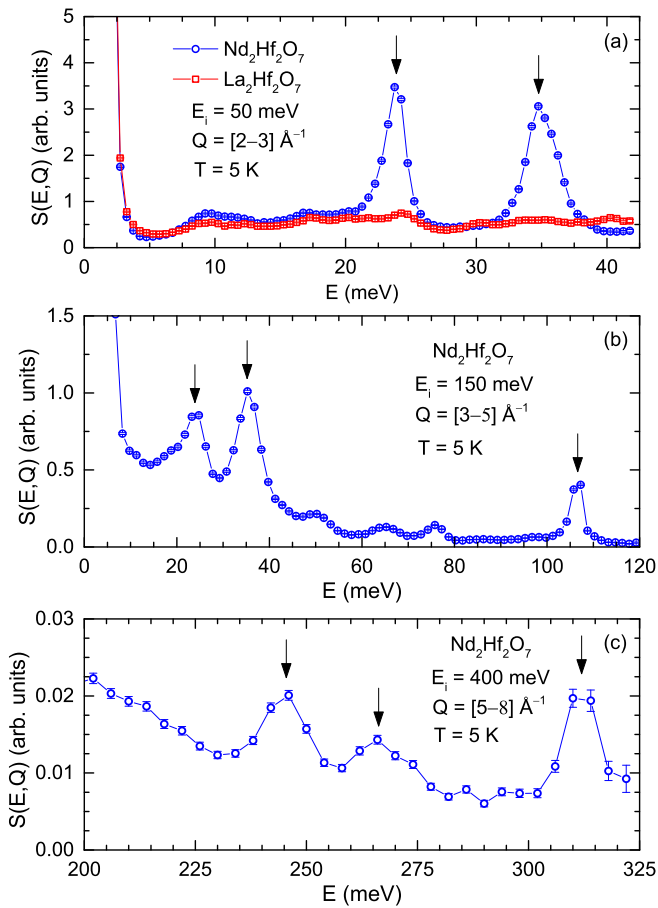}
\caption {(Color online) $Q$-integrated inelastic scattering intensity $S(E,Q)$ versus energy transfer $E$ of Nd$_2$Hf$_2$O$_7$ for (a) $E_i=50$~meV, momentum $Q$ range [2--3]~{\AA $^{-1}$}, (b) $E_i=150$~meV, $Q$ range [3--5]~{\AA $^{-1}$}, and (c) $E_i=400$~meV, $Q$ range [5--8]~{\AA $^{-1}$} at 5 K\@. The arrows mark the crystal electric field (CEF) excitations. The CEF excitations in (a) and (b) arise from the transitions from the ground state multiplet $^4I_{9/2}$ of Nd$^{3+}$ and in (c) from the transitions to the first excited state multiplet $^4I_{11/2}$.}
\label{fig:INS2}
\end{figure}

The inelastic neutron scattering responses from Nd$_2$Hf$_2$O$_7$ and La$_2$Hf$_2$O$_7$ are shown in Fig.~\ref{fig:INS1} as color-coded contour maps depicting the energy transfer $E$ versus wave vector $Q$ for neutrons of incident energies $E_i=50$~meV, 150~meV and 400~meV at 5~K\@. The La$_2$Hf$_2$O$_7$ being nonmagnetic shows only scattering of phonons whose intensity increases with increasing $Q$. The Nd$_2$Hf$_2$O$_7$, on the other hand, in addition to these phonon scattering, also shows three strong dispersionless excitations around 23.7~meV, 34.7~meV, and 106.5~meV and three weaker excitations around 245.7~meV, 265.9~meV and 311.9~meV which are quite clear for the low-$Q$ region of the contour plots. These excitations are marked with arrows in Fig.~\ref{fig:INS1}. The intensities of these low-$Q$ excitations decrease with increasing $Q$ which suggests that they have magnetic origin due to the crystal field excitations from  Nd$^{3+}$. The $Q$-integrated one-dimensional energy cuts of the INS response are shown in Fig.~\ref{fig:INS2} which illustrate the CEF excitations more clearly. No additional magnetic excitations were resolvable in the INS spectra measured with $E_i=700$~meV (not shown).  

Typically for Nd$^{3+}$ ion the transitions from the ground state (GS) multiplet $^4I_{9/2}$ is below 200~meV, therefore the excitations 23.7~meV, 34.7~meV, and 106.5~meV are understood to arise from the transitions within the GS multiplet, whereas the excitations 245.7~meV, 265.9~meV and 311.9~meV [Fig.~\ref{fig:INS2}(c)] are assigned to the transitions from the first excited multiplet $^4I_{11/2}$. For Nd$^{3+}$ $J = 9/2$, therefore the ($2J+1 = 10$)-fold degenerate GS multiplet $^4I_{9/2}$, when subject to CEF created by the eight neighboring oxygen ions in the cubic pyrochlore structure (with $D_{3d}$ symmetry) of Nd$_2$Hf$_2$O$_7$, should split into five doublets of $|\pm m_J\rangle$ type. Accordingly, one would expect four excitations for the transitions from the ground state doublet to the four excited doublet states. However, we see only three excitations at 23.7~meV, 34.7~meV, and 106.5~meV. The peak at ~34.7 meV in Fig.~\ref{fig:INS2}(a) appears broader than the peak at 23.7~meV, indicating for the presence of two unresolved excitations from two closely situated CEF levels in the vicinity of 34.7~meV corresponding to the so-called quasi-quartet state as inferred from the analysis of magnetic heat capacity data \cite{Anand2015}. The INS spectra of Nd$_2$Zr$_2$O$_7$ was also found to show similar unresolved excitations near 35.0~meV \cite{Xu2015,Lhotel2015}. We tried to fit the 34.7~meV with two Lorentzian peaks which yielded the two possible peaks at 34.7 and 35.8~meV.

\begin{figure}
\includegraphics[width=\columnwidth, keepaspectratio]{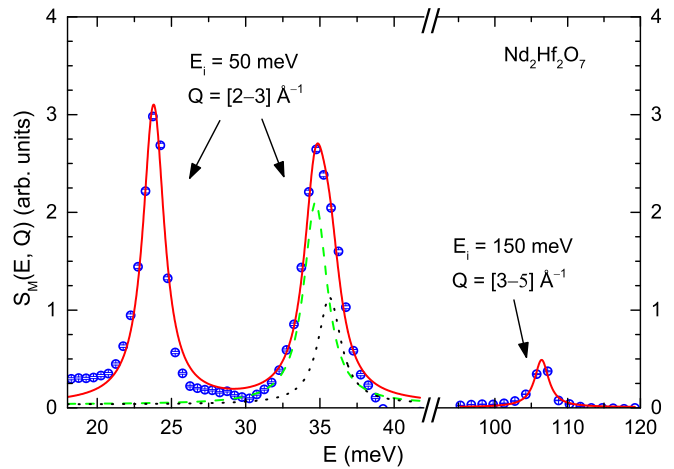}
\caption {(Color online) $Q$-integrated inelastic magnetic scattering intensity $S_{\rm M}(E,\omega)$ versus energy transfer $E$ for Nd$_2$Hf$_2$O$_7$. The data for the three lower crystal field levels are from the dataset with incident neutron energy $E_i=50$~meV integrated over the momentum $Q$ range [2--3]~{\AA $^{-1}$}. The data for the crystal field excitation at 106.4~meV are from the dataset with $E_i=150$~meV, integrated $Q$ range [3--5]~{\AA $^{-1}$}. The solid lines are the fits of the data according to the crystal field model discussed in the text. The dashed and dotted lines show two closely spaced peaks near 34.7~meV and 35.7~meV. }
\label{fig:INS3}
\end{figure}

We analyze the INS data by a model based on the crystal electric field. Further we use tensor operator formalism (instead of Stevens' operator)  which allows us to account for the mixing of the GS multiplet with the higher multiplets. In the tensor operator formalism the CEF Hamiltonian for the fcc pyrochlore structure having $D_{3d}$ symmetry (point symmetry $\bar{3}m$) with local cubic $\langle$111$\rangle$ direction along the $z$ axis is given by  \cite{cfbook}
\begin{equation}
\begin{split}
H_{\rm{CEF}}= & B_0^2C_0^2+B_0^4C_0^4+B_3^4(C_{-3}^4+C_3^4) +B_0^6C_0^6\\
             & +B_3^6(C_{-3}^6+C_3^6)+B_6^6(C_{-6}^6+C_6^6).
\end{split}
\end{equation}
The $B_q^k$ are the crystal field parameters and $C_q^k$ the tensor operators in Wybourne notation \cite{cfbook}. Following the approach we used for Nd$_2$Zr$_2$O$_7$ \cite{Xu2015}, we employ a set of 108 intermediate coupling basis states which account for the basis states from 12 multiplets below 2.24~eV. The results of our analysis of the magnetic scattering data (obtained after subtracting the phonon contribution using the La$_2$Hf$_2$O$_7$ INS data) using the software SPECTRE \cite{SPECTRE} are summarized in Table~\ref{tab:cf}. The fits of magnetic excitations with a Lorentzian shape peak function is shown in Fig.~\ref{fig:INS3}. For least-squares fitting of the observed excitation energies and relative intensities the starting $B_q^k$ parameters were taken to be equivalent to those of Nd$_2$Zr$_2$O$_7$ \cite{Xu2015}.

\begin{table}
\caption{\label{tab:cf} Observed  and calculated  crystal-field transition energies ($E$) and integrated intensities ($I$) within the ground state multiplet $^4I_{9/2}$ of N$_2$Hf$_2$O$_7$ at 5~K\@. The $I$ is relative with respect to the highest peak observed. The $B_q^k$ parameters obtained from the analysis are: $B^2_0 = 49.8$~meV, $B^4_0 = 419.0 $~meV, $B^4_3 = 121.2 $~meV, $B^6_0 = 142.9 $~meV, $B^6_3 = -94.5 $~meV, and $B^6_6 = 140.9$~meV.} 
\begin{ruledtabular}
\begin{tabular}{ccccc}
Levels & $E_{obs}$ (meV)& $E_{cal}$ (meV)&$I_{obs}$&$I_{cal}$\\
\hline
$\Gamma_{56}^+$ &0 & 0 & - & 2.5\\
$\Gamma_4^+$ & $23.7(3)$ & 23.84 & $0.8(1)$ & 0.78\\
$\Gamma_{56}^+$ & $34.7(3)$ & 34.67 &  1.0 &1.0\\
$\Gamma_4^+$ & $35.8(3)$ & 35.65 &0.5(1) & 0.55\\
$\Gamma_4^+$ & $106.5(5)$ & 106.57 & $0.6(1)$ & 0.74\\
\end{tabular}
\end{ruledtabular}
\end{table}

\begin{figure}
\includegraphics[width=\columnwidth, keepaspectratio]{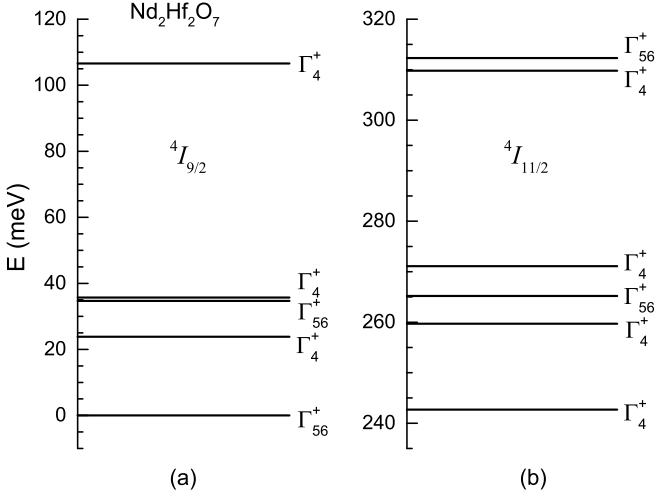}
\caption {(Color online) Crystal field energy schemes for (a) the ground-state multiplet $^4I_{9/2}$ and (b) the first excited multiplet $^4I_{11/2}$ corresponding to the crystal field parameters obtained from the analysis of INS data. $\Gamma$ shows the irreducible representation that the corresponding CEF state transforms as.}
\label{fig:INS4}
\end{figure}

The least square fits of the INS data yielded the $B_q^k$ parameters $B^2_0 = 49.8$~meV, $B^4_0 = 419.0 $~meV, $B^4_3 = 121.2 $~meV, $B^6_0 = 142.9 $~meV, $B^6_3 = -94.5 $~meV, and $B^6_6 = 140.9$~meV which correspond to five doublets at 0, 23.8~meV, 34.7~meV, 35.7~meV and 106.6~meV. The CEF energy level scheme of the GS multiplet $^4I_{9/2}$ obtained from the analysis of INS data is shown in Fig.~\ref{fig:INS4}. Also shown are the calculated energy levels for the transitions to the first excited multiplet $^4I_{11/2}$ which are in good agreement with the experimental results [Fig.~\ref{fig:INS2}(c)]. 

In order to check if the ground state doublet of Nd$^{3+}$ in N$_2$Hf$_2$O$_7$ is consistent with the dipolar-octupolar nature we first convert the $B_q^k$ parameters into the Steven's formalism by using the relation $D^q_k=\Lambda \lambda^q_k B^q_k$ ($\Lambda=\alpha_J,\ \beta_J \ {\rm and}\  \gamma_J$ listed in Ref.~\cite{hutchings} and $\lambda^q_k$ listed in Ref.~\cite{bookliu}). An essential condition for the Kramers doublet GS of Nd$^{3+}$ ($J = 9/2$) to be compatible with dipolar-octupolar character is that the parameter $D^2_0$ be negative and dominate over other terms \cite{Huang2014}. The transformation of $B_q^k$ parameters into the Steven's formalism gives $D^2_0 = - 0.160 $~meV, $D^4_0 = - 0.0153 $~meV, $D^4_3 =0.105$~meV, $D^6_0 = -0.00034 $~meV, $D^6_3 = -0.0046. $~meV, and $D^6_6 = -0.005$~meV.  We see that the $D^2_0$ is indeed negative and dominating which confirms that the Kramers doublet ground state of Nd$^{3+}$ is consistent with the dipolar-octupolar character. 

\begin{figure}
\includegraphics[width=\columnwidth, keepaspectratio]{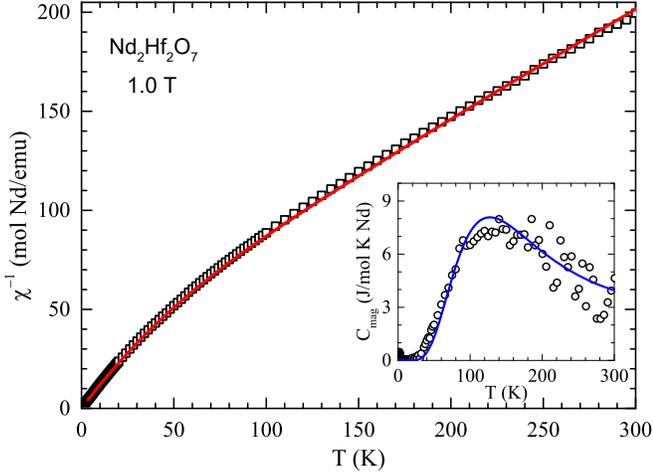}
\caption {(Color online) Inverse magnetic susceptibility $\chi^{-1}$ versus temperature $T$ of polycrystalline Nd$_2$Hf$_2$O$_7$ measured in a field of 1.0~T\@. The red solid line is the inverse crystal field susceptibility $\chi_{\rm CEF}^{-1}(T)$ corresponding to the crystal field parameters obtained from the analysis of INS data. Inset: Magnetic contribution to heat capacity $C_{\rm mag}(T)$ \cite{Anand2015}. The blue solid curve is the heat capacity calculated from the crystal field $C_{\rm CEF}(T)$.}
\label{fig:INS5}
\end{figure}

Further we find significant mixing of $m_J$ terms within the GS multiplet $^4I_{9/2}$. The analysis of INS data yielded the wavefunction for the Kramers doublet ground state to be
\begin{subequations}
\begin{equation}
\begin{split}
\Gamma_{56}^+  = & \ 0.903|^4I_{9/2},\pm 9/2 \rangle + 0.334|^4I_{9/2}, \mp 3/2 \rangle \\
                   & \mp 0.232|^4I_{9/2}, \pm 3/2 \rangle  \mp 0.111|^4I_{11/2},\pm 9/2 \rangle \\                   
 			& + 0.045|^4I_{13/2},\pm 9/2 \rangle \\ 
\end{split}
\end{equation}
and the wavefunction of the first excited doublet is found to be
\begin{equation}
\begin{split}
\Gamma_4^+    =  &   \ 0.139|^4I_{9/2}, \pm 7/2 \rangle + 0.720|^4I_{9/2},\mp 5/2 \rangle  \\ 
               &  \mp 0.671|^4I_{9/2}, \pm 1/2 \rangle \pm 0.053|^4I_{11/2},\pm 7/2 \rangle \\
               &   \pm 0.052|^4I_{11/2},\pm 5/2 \rangle. 
\end{split}
\end{equation}
\label{eq:wavefunctions}
\end{subequations}
The ground state wavefunction [Eq.~(\ref{eq:wavefunctions}a)] clearly shows the mixing of $|^4I_{9/2},\pm 9/2 \rangle$ with $|^4I_{9/2},m_J\neq\pm9/2 \rangle$ terms as well as with excited state multiplets $^4I_{11/2}$ and $^4I_{13/2}$. The mixing of $m_J$ terms and higher state multiplets can be held responsible for the observation of reduced ordered moment of Nd$^{3+}$ \cite{Anand2015}. We estimate the magnetic moment from the ground state wavefunction in Eq.~(\ref{eq:wavefunctions}a) which turns out to be $2.53\,\mu_{\rm B}$, consistent with the value of effective moment obtained from the magnetic susceptibility data ($\approx 2.45\,\mu_{\rm B}$) \cite{Anand2015}. Furthermore we find the $g$-parameter $g_{\rm zz}\approx 5.1$ and $g_{\bot}=0$ which are again close to the values obtained from the magnetic data \cite{Anand2015}. The values of ground state moment and $g$-parameters reflect the Ising anisotropy in Nd$_2$Hf$_2$O$_7$. The large splitting between the ground state doublet and the first excited state doublet suggests that the Ising anisotropy is quite strong. 

Next we estimate the crystal field susceptibility $\chi_{\rm CEF}(T)$ and heat capacity $C_{\rm CEF}(T)$ using the CEF parameters listed in Table~\ref{tab:cf}. The calculated $\chi_{\rm CEF}(T)$ is shown in Fig.~\ref{fig:INS5} (red solid line) along with the experimental $\chi(T)$ data plotted as their inverse, and we see a very reasonable agreement between $\chi_{\rm CEF}(T)$ and $\chi(T)$. The ratio $\chi_\parallel/\chi_\perp $ of the anisotropic CEF suceptibility $\chi_\parallel$ (parallel to $\langle 111 \rangle$) and $\chi_\perp$ (perpendicular to $\langle 111 \rangle$) is found to be $ \sim 67 $ at 10~K and $ \sim 191 $ at 3.5~K\@. For comparison, the value of $\chi_\parallel/\chi_\perp $ for Dy$_2$Ti$_2$O$_7$ is $ \sim 300 $ at 10~K \cite{Princep2015}. For hafnate pyrochlore Pr$_2$Hf$_2$O$_7$, $\chi_\parallel/\chi_\perp \sim 45$ at 10~K  \cite{Anand2016}. The $C_{\rm CEF}(T)$ calculated from the CEF parameters is shown in the inset of Fig.~\ref{fig:INS5} (solid blue curve) which again is in very reasonable aggreement with the magnetic heat capacity $C_{\rm mag}(T)$ data (see Ref.~\cite{Anand2015} for a description of the measurement and phonon subtraction of heat capacity). The good agreement of $\chi_{\rm CEF}(T)$ and $C_{\rm CEF}(T)$ with the experimental data supports the deduced CEF parameters and energy level scheme.

\section{\label{Conclusion} Conclusions}

We have investigated the spin dynamics of all-in/all-out antiferromagnet Nd$_2$Hf$_2$O$_7$ using the muon spin relaxation technique and determined the crystal field states through the inelastic neutron scattering measurements. The INS data show three CEF excitations near 23.7~meV, 34.7~meV and 106.5~meV due to transitions from the ground state doublet. The excitation near 34.7~meV is rather broad and comprises of two unresolved excitations. The analysis of INS data suggests the five doublets of Nd$^{3+}$ to be situated at 0, 23.8~meV, 34.7~meV, 35.7~meV and 106.6~meV.  The anisotropic $g$-parameters obtained from the CEF-split Kramers doublet ground state wavefunction along with the large CEF splitting between the ground state and the first excited doublet provide evidence for strong local $\langle111\rangle$ Ising anisotropy as previously noted from the analysis of magnetic susceptibility and isothermal magnetization \cite{Anand2015}. The CEF parameters obtained from the analysis of INS data reveal the dipolar-octupolar nature of Kramers doublet ground state of Nd$^{3+}$ ($J=9/2$) moments, making Nd$_2$Hf$_2$O$_7$ a candidate compound for dipolar and octupolar spin ice phases \cite{Huang2014}. 

The zero field $\mu$SR data confirm the occurence of long range magnetic ordering below $T_{\rm N} = 0.55$~K. Further, the ZF $\mu$SR also show evidence for the presence of persistent dynamic spin fluctuations deep inside the ordered state, which is manifested by novanishing depolarization rates in the Lorentzian channel in the limit of $T\rightarrow0$. The longitudinal field $\mu$SR data show the dramatic effect of external magnetic field on the fast relaxing component of the Lorentzian channel. From the field dependence of this depolarization rate we estimate a correlation time of about 600~ns for the spin fluctuations. The reduced ordered moment can thus be attributed to the presence of persistent dynamic fluctuations very likely on account of mixing of the $m_J$ terms and higher state multiplets as well as the octupolar tensor component of Kramers doublet ground state of Nd$^{3+}$ ($J=9/2$). 
 
The all-in/all-out antiferromagnetic ordering, the CEF level scheme, strong Ising anisotropy, dipolar-octupolar nature of Kramers doublet ground state of Nd$^{3+}$, and persistent dynamic fluctuations in the ordered state are the common features that Nd$_2$Hf$_2$O$_7$ shares with Nd$_2$Zr$_2$O$_7$ \cite{Xu2015,Lhotel2015, Xu2016}. We thus see that apart from the difference in detecting the long range ordered state through $\mu$SR, the ground state properties of Nd$_2$Hf$_2$O$_7$ are very similar to those of Nd$_2$Zr$_2$O$_7$. Therefore in view of the recent observation of magnetic fragmentation and hence spin-ice behavior in Nd$_2$Zr$_2$O$_7$ \cite{Petit2016} one would naively expect a similar behavior in Nd$_2$Hf$_2$O$_7$, which, however, remains to be verified experimentally. The Pr analog of Nd$_2$Hf$_2$O$_7$, Pr$_2$Hf$_2$O$_7$ has recently been found to present signatures of quantum spin-ice behavior without any evidence of long range magnetic ordering down to 90~mK \cite{Anand2016,Sibille2016}. Further investigations to understand the possible attributes due to dipolar-octupolar character and/or spin fragmentation in Nd$_2$Hf$_2$O$_7$ are underway.

\acknowledgements
We thank J. Xu for his help in analysis of INS data and fruitful discussions. Helpful discussions on $\mu$SR data analysis with Peter Baker and Francis Pratt are gratefully acknowledged. We acknowledge Helmholtz Gemeinschaft for funding via the Helmholtz Virtual Institute (Project No. VH-VI-521). The research at ORNL's Spallation Neutron Source was sponsored by the Scientific User Facilities Division, Office of Basic Energy Sciences, US Department of Energy.

\end{document}